\documentclass[reprint,superscriptaddress,amsmath,amssymb,aps,prb,showpacs]{revtex4-1}
\usepackage[]{graphicx}
\usepackage{dcolumn}
\usepackage{bm}
\usepackage{hyperref}

\usepackage{todonotes}

\begin{document}

\title{Correlated Diffuse X-ray Scattering from Periodically Nano-Structured Surfaces}

\author{V. Soltwisch}
\email{Victor.Soltwisch@ptb.de}
\affiliation{Physikalisch-Technische Bundesanstalt (PTB), 
Abbestr. 2-12, 10587 Berlin, Germany}

\author{A. Haase}
\affiliation{Physikalisch-Technische Bundesanstalt (PTB), 
Abbestr. 2-12, 10587 Berlin, Germany}

\author{J. Wernecke}
\affiliation{Physikalisch-Technische Bundesanstalt (PTB), 
Abbestr. 2-12, 10587 Berlin, Germany}

\author{J. Probst}
\affiliation{Helmholtz-Zentrum Berlin (HZB), Albert-Einstein-Str. 15,
12489 Berlin, Germany}

\author{M. Schoengen}
\affiliation{Helmholtz-Zentrum Berlin (HZB), Albert-Einstein-Str. 15,
12489 Berlin, Germany}

\author{S. Burger}
\affiliation{Zuse Institute Berlin (ZIB), Takustra\ss{}e 7,
14195 Berlin, Germany}
\author{M. Krumrey}
\affiliation{Physikalisch-Technische Bundesanstalt (PTB), 
Abbestr. 2-12, 10587 Berlin, Germany}

\author{F. Scholze}
\affiliation{Physikalisch-Technische Bundesanstalt (PTB), 
Abbestr. 2-12, 10587 Berlin, Germany}

\date{\today}

\begin{abstract}
Laterally periodic nanostructures were investigated with grazing incidence small angle X-ray scattering. To support an improved reconstruction of nanostructured surface geometries, we investigated the origin of the contributions to the diffuse scattering pattern which is correlated to the surface roughness. 
Resonant diffuse scattering leads to a palm-like structure of intensity sheets. Dynamic scattering generates the so-called Yoneda band caused by a resonant scatter enhancement at the critical angle of total reflection and higher-order Yoneda bands originating from a subsequent diffraction of the Yoneda enhanced scattering at the grating.
Our explanations are supported by modelling using a solver for the time-harmonic Maxwell's equations based on the finite-element method.

\end{abstract}

\maketitle
\section{introduction}
X-ray scattering is established as a common technique in nano science. Among the diverse nano-sized objects presently under investigation in natural and life science~\cite{Weinhausen:2014,Auguie:2008,Narayanan:2004},
the continuously shrinking patterns in the semiconductor industry are at the forefront regarding their requirements for size reproducibility and regularity~\cite{markov_limits_2014}.
The latter also differ from many other applications which benefit from the single particle diffraction measurements now feasible with free-electron laser sources~\cite{
Sun_2012fk,chapman_femtosecond_2011,Barke:2015uq}.
Measurements on the length scale of several nm are basically challenged by the atomic granularity of matter and by structures which are not perfect and sharply defined anymore. Due to the large photon beam footprint as compared to the nm pattern size, grazing incidence small angle X-ray scattering (GISAXS)~\cite{guinier_small-angle_1955,levine_grazing-incidence_1989} directly yields statistical information for a large structured area on fluctuations like structure roughness. It is a powerful tool for the analysis of the morphology and distribution of nanoparticles on surfaces or buried particles~\cite{PhysRevLett.94.145504,jiang_waveguide-enhanced_2011}. As compared to SAXS in transmission geometry~\cite{wang_saxs_2007,Jones_2006}, GISAXS also allows to study structured surfaces of thick, non-homogeneous substrates.

Recent GISAXS studies on nanoscale line gratings ~\cite{hofmann_grazing_2009,rueda_grazing-incidence_2012,wernecke_direct_2012-1,gollmer_fabrication_2014} mainly focus on the grating diffraction and deliver no explanation for the diffuse scattering background.
Corresponding theoretical calculations for small angle X-ray scattering in the past were mostly done with the distorted wave Born approximation (DWBA)~\cite{Babonneau:hx5104,Lazzari:vi0158,renaud_probing_2009-1,rueda_grazing-incidence_2012,rauscher_small-angle_1995-1,jiang_waveguide-enhanced_2011} including an analytic expression of the line shape.
Arbitrarily shaped structures must be discretized and require numerical solvers \cite{Chourou:nb5076}. For periodic structures, e.g.~gratings, it is well established to model the light scattering by numerically solving the time-harmonic Maxwell's equations with a higher-order finite-element method ~\cite{pomplun_adaptive_2007,Kato:12}. If the periodic structures are invariant in one dimension (along the grating lines), the computational domain can be reduced to a two dimensional problem which decreases the computational effort significantly and allows to calculate also rather large domains as compared to the incident wavelength. This helps to apply the method to short-wavelengths like in X-ray scattering. The computation yields a near field solution for the electric field in the structured volume. This can be post-processed to obtain the far field solution, i.e.~the scattering intensities, or other near field parameters like the electric field strength along a certain path, e.g.~the vacuum-material interface.\\

  \begin{figure}[bp]
 	\centering
     \includegraphics[width=0.48\textwidth]{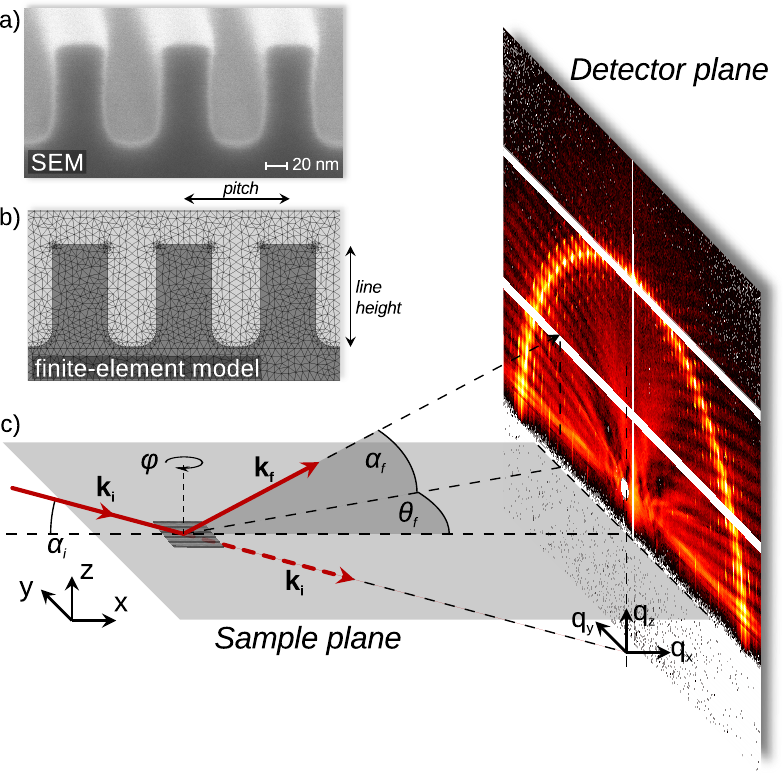}
 \caption{(a): SEM image of a silicon grating sample similar to grating No.~5 (see table in Fig.~\ref{figure_yoneda_waveguide}). (b): Finite-element model employed for the simulations of above grating sample. (c): Scattering of a collimated X-ray beam under grazing incidence from a lamellar grating oriented parallel ($\varphi=0^\circ$) with respect to the incident beam.}
 \label{figure_sketch}
 \end{figure}
 
The scope of this investigation is to determine the kinematic and dynamic scattering processes which cause the diffuse background and relate them to geometrical parameters of the scattering surface in order to obtain more detailed information on nano-structured surface geometries. 
Yoneda enhanced surface scattering~\cite{vineyard_grazing-incidence_1982,yoneda_anomalous_1963} and subsequent diffraction at a periodic nanostructure give rise to the formation of higher-order Yoneda bands in the diffuse scattering background. Meanwhile, the origin of the Yoneda bands and additional waveguide effects can be explained with a grating effective layer.
\section{Experimental Details}
We investigated lamellar silicon gratings 
as a prototype specimen for state-of-the-art integrated electronic elements. The grating structures were prepared by electron beam lithography and reactive ion etching on a silicon wafer. The grating area measures 1 mm by 15 mm with the lines oriented parallel to the long edge. This accommodates the elongated footprint of the beam in GISAXS geometry. The five samples discussed here are listed in the table below Fig.~\ref{figure_yoneda_waveguide}. 

The experiments were conducted at the PTB's four crystal monochromator beamline~\cite{krumrey_high-accuracy_2001} at the electron storage ring BESSY II, which is equipped with an in-vacuum Pilatus 1M area detector~\cite{wernecke_characterization_2014}. 
The GISAXS scheme is illustrated in Fig.~\ref{figure_sketch}c: A monochromatic X-ray beam with a wavevector $\vec{k}_i$ impinges on the sample surface at a grazing incidence angle $\alpha_i$ and an azimutal angle $\theta_i=0^\circ$. The elastically scattered wavevector $\vec{k}_f$ propagates along the exit angle $\alpha_f$ and the azimuthal angle $\theta_f$.
\section{Diffuse scatter contributions from structured surfaces}
\subsection{Yoneda band - Waveguide effects}
The regular structure of a grating leads to discrete diffraction orders on a cone with an opening angle equal to the incidence angle $\alpha_i$. Roughness and imperfections of the grating give rise to satellite diffraction orders, which result from the fabrication process and lead to different diffuse scattering contributions.
\begin{figure}[tbp]
	\centering
    \includegraphics[width=0.43\textwidth]{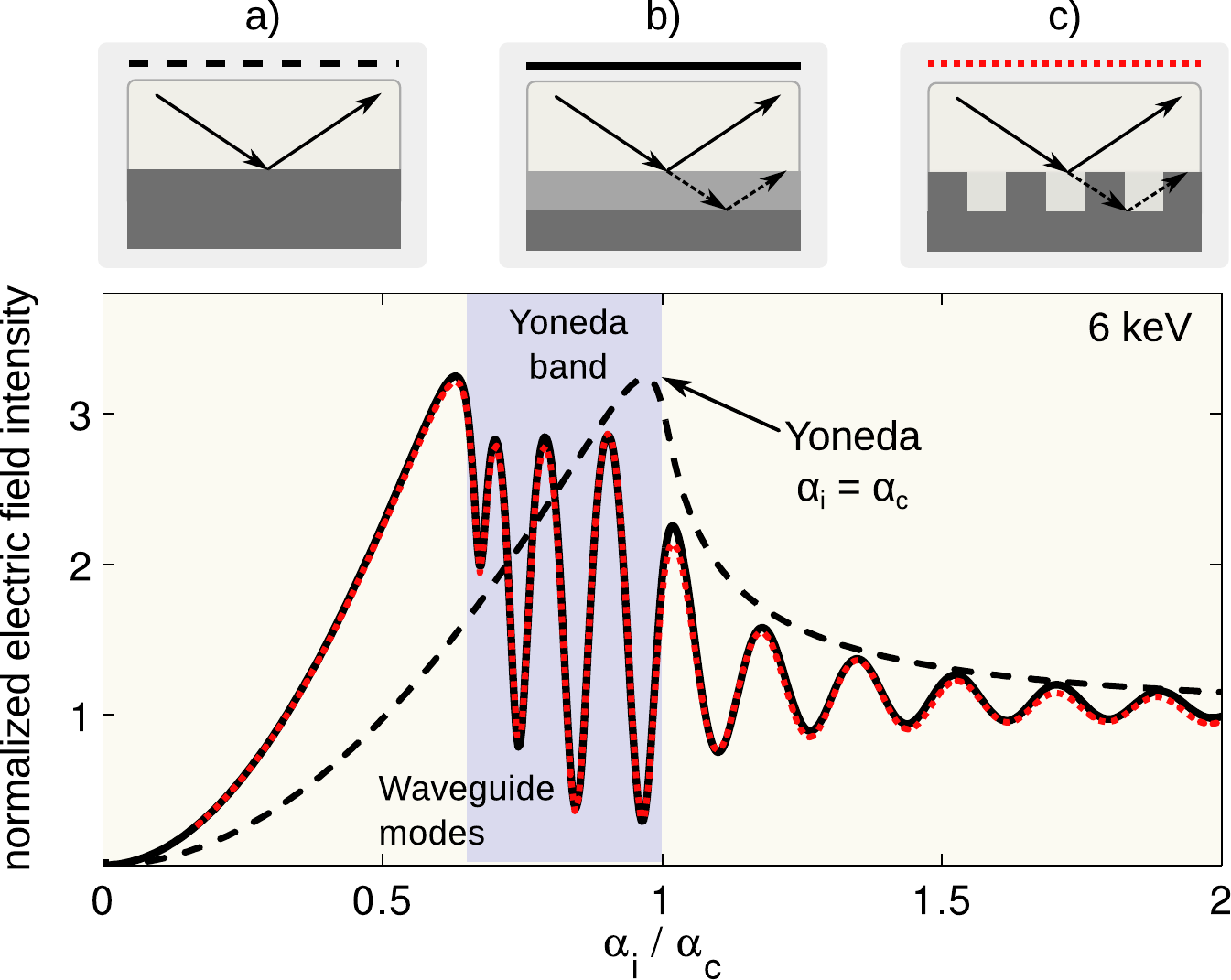}
\caption{
Calculated electric field intensities at the vacuum interface for incident angles $\alpha_i$ around the critical angle of silicon.
(a): The dashed black line represents the surface field enhancement at the critical angle for a flat Si-substrate. (b): The substrate was capped with an additional layer with reduced mass density (42\%) and a thickness of 100 nm.  (c): The field intensity for a grating with rectangular lines at oblique azimuthal incidence ($\varphi = 2^{\circ}$). 
}
\label{figure_fields}
\end{figure}
\begin{figure}[htbp]
	\centering
  \includegraphics[width=0.489\textwidth]{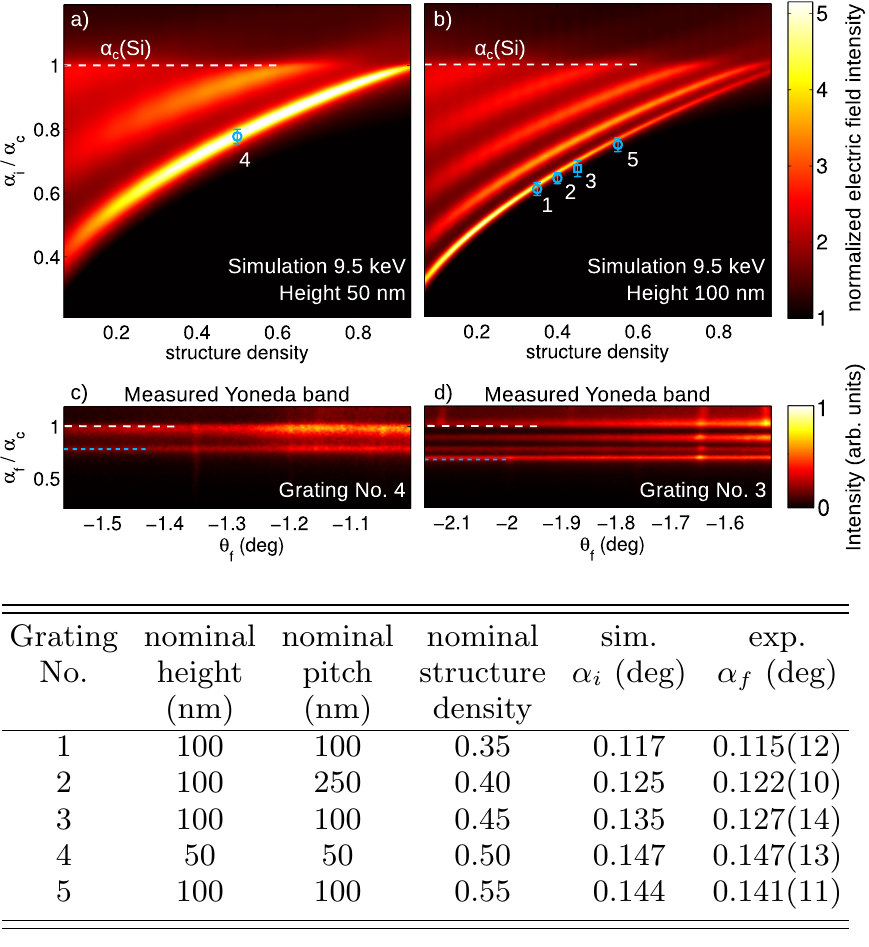}
 \caption{
The electric field intensity at the vacuum interface normalized to the incident plane wave calculated for $\alpha_i$ below the critical angle of silicon for a rectangular grating with a line height of 50 nm (a) and 100 nm (b) as a function of structure density. The angular scale is normalized to $\alpha_c$. Blue circles in both figures mark the experimentally determined position $\alpha_f$ of the lower Yoneda band boundary. The numerical values for the simulated angles $\alpha_i$ and measured $\alpha_f$ are summarized in the table. (c,d): Measured Yoneda band characteristic obtained from grating No.~4 and grating No.~3.
}
 \label{figure_yoneda_waveguide}
\end{figure}
If the exit angle $\alpha_f$ is equal to the critical angle of total reflection $\alpha_c$, the diffuse scattering from a rough surface is resonantly enhanced.
This resonant enhancement is often called the Yoneda or Vineyard peak~\cite{vineyard_grazing-incidence_1982,yoneda_anomalous_1963} which appears here as a bright horizontal line of enhanced diffuse scattering intensity. 
The mechanism behind this enhancement is visualized in Fig.~\ref{figure_fields}
by showing the electric field intensity at the surface for incident angles around the critical angle (a). The field intensity, and thus the potential scattering from the surface, reaches its maximum at the critical angle.

If the surface is capped with a thin layer (b), the Yoneda line splits up into a band with pronounced oscillations between the critical angles of the substrate and the capping layer caused by the formation of an X-ray standing wave inside the capping layer~\cite{wang_resonance-enhanced_1992}. The incident wave is coupled to these waveguide modes~\cite{feng_x-ray_1993}. This effect has been used for depth profiling of thin films or buried nanostructures~\cite{jiang_waveguide-enhanced_2011,PhysRevLett.94.145504,Babonneau_PhysRevB}.
Here, we observed a similar Yoneda band in the GISAXS pattern from laterally structured surfaces instead of a simple layer system.
Electromagnetic field computations for a rectangular grating (c) and oblique incident angles $\alpha_i$
demonstrate that the local field enhancement above the grating behaves almost identical to that of a top layer with reduced density.
For diffraction efficiency measurements this behavior is well known~\cite{lee_nanoimprint_2005,tolan_x-ray_1995} from scattering geometries perpendicular to the grating lines ($\varphi=90^\circ$).
\begin{figure}[tb]

  \includegraphics[width=0.5\textwidth]{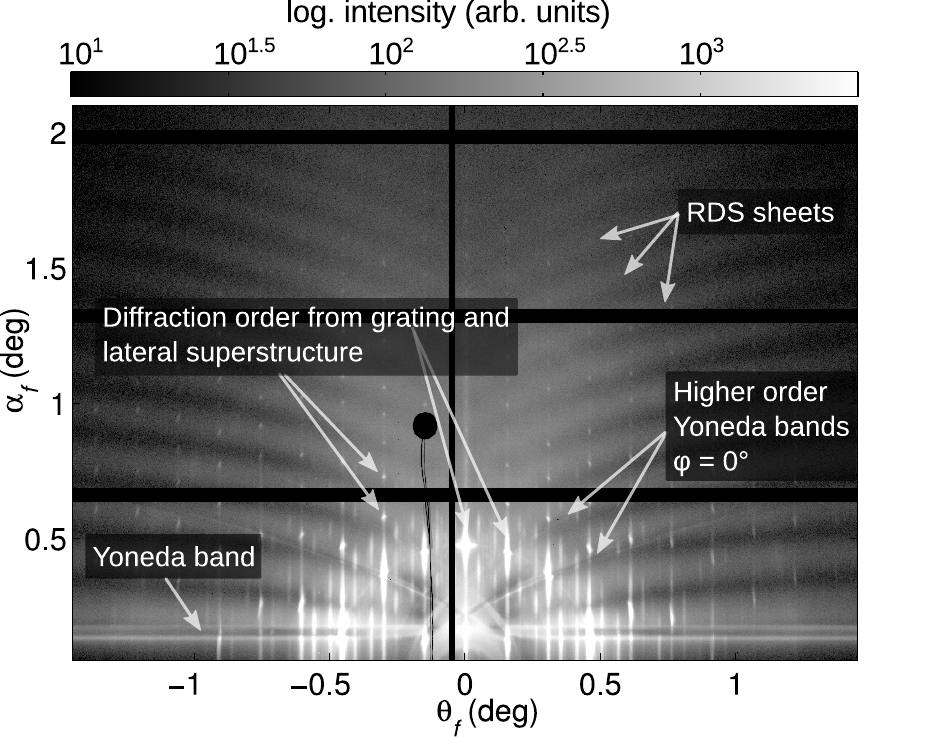}
\caption{GISAXS pattern from grating No.~4 obtained at \mbox{9.5 keV} and $\alpha_i=0.5^\circ$, $\varphi=0^\circ$ with logarithmic gray scale. 
The range of the scale was adjusted to optimize the visibility of the diffuse scattering background.}
\label{figure_explain_gisaxs}
\end{figure}

Here, we investigated the Yoneda band characteristics for several lamellar silicon gratings with different geometrical dimensions (see Table in Fig.~\ref{figure_yoneda_waveguide}) close to the parallel orientation of the lines with respect to the incoming beam ($\varphi=0^\circ$). 
The number of waveguide modes depends on the structure height and density, as illustrated in Fig.~\ref{figure_yoneda_waveguide}. The electric field at the vacuum interface, following the line profile into the grating grooves, was computed for an azimuthally rotated rectangular grating ($\varphi = 1^{\circ}$) and incident angles $\alpha_i$ around the critical angle of silicon. For the grating with 50 nm nominal height and a 1:1 duty cycle, measurement and electric field computation consistently yield a splitting of the Yoneda band, illustrated in Fig.~\ref{figure_yoneda_waveguide}c. The gratings with 100 nm nominal height show pronounced oscillations inside the Yoneda band very similar to waveguide modes (Fig.~\ref{figure_yoneda_waveguide}b). Variation of the grating linewidths reveals the high sensitivity of the Yoneda band for structure density, i.e.~the line-to-pitch ratio, as well as height variations. Vice versa, experimental values for the linewidth with respect to the pitch can be obtained by matching the calculated and measured (Fig.~\ref{figure_yoneda_waveguide} blue circles) lower band boundaries. The sharpness of the Yoneda bands for the individual samples varied, resulting in different error margins as shown in Fig.~\ref{figure_yoneda_waveguide}.
\subsection{Higher-order Yoneda scattering}
\begin{figure}[b]
	\centering
  \includegraphics[width=0.46\textwidth]{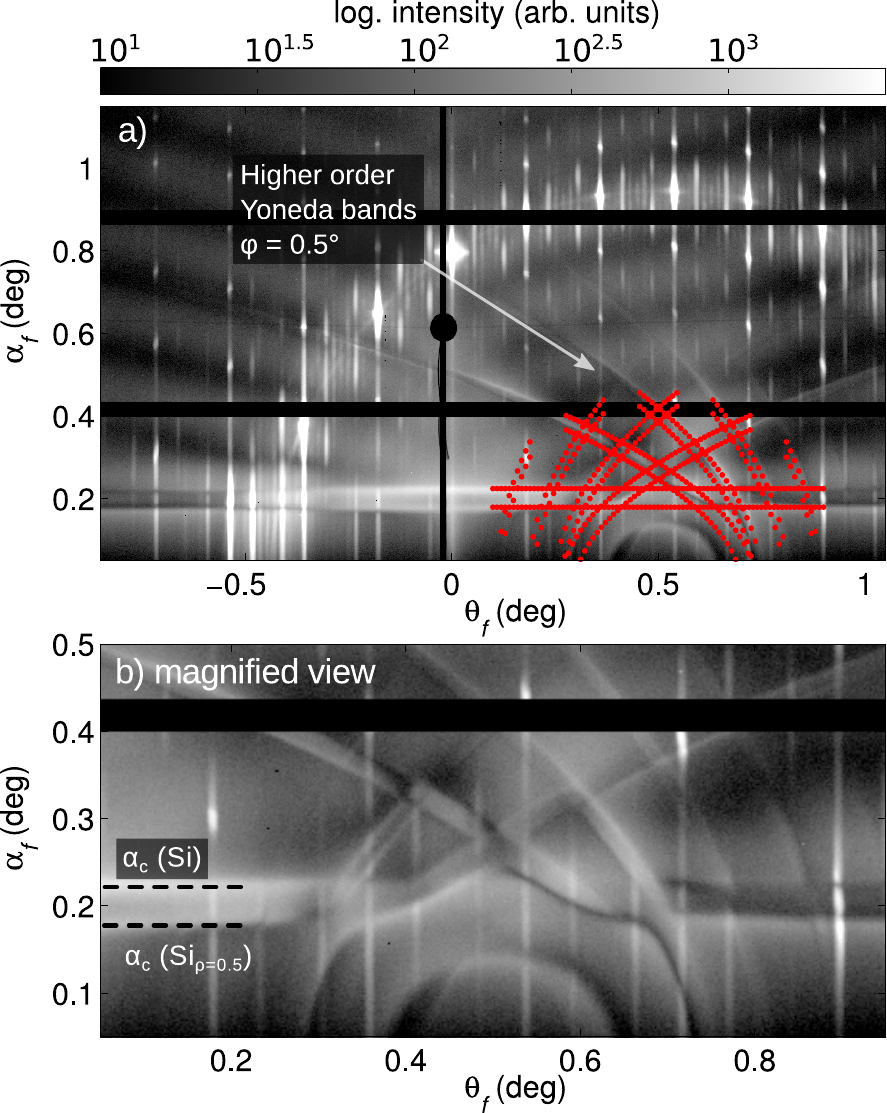}
\caption{(a): The central area under an azimuthal rotation ($E=8$ keV, $\alpha_i=0.8^\circ$, $\varphi=0.5^\circ$) of the grating No.~4. The diffraction cone and higher order Yoneda bands follow this rotation, in contrast to the RDS sheets. Red dots illustrate the expected reciprocal space locations for the higher orders of the Yoneda band.
(b): Magnified view of figure (a) around the higher-order Yoneda bands.}
\label{figure_explain_gisaxs2}
\end{figure}
The Yoneda scattering process gives rise to an additional structure in the diffuse scattering background.
In contrast to a homogeneous layer, the grating causes also higher ordering of the Yoneda band, appearing as bent sharp lines originating from the sample horizon and crossing the Yoneda band (Fig.~\ref{figure_explain_gisaxs}).
\begin{figure*}[tbp]
	\centering
  \includegraphics[width=0.95\textwidth]{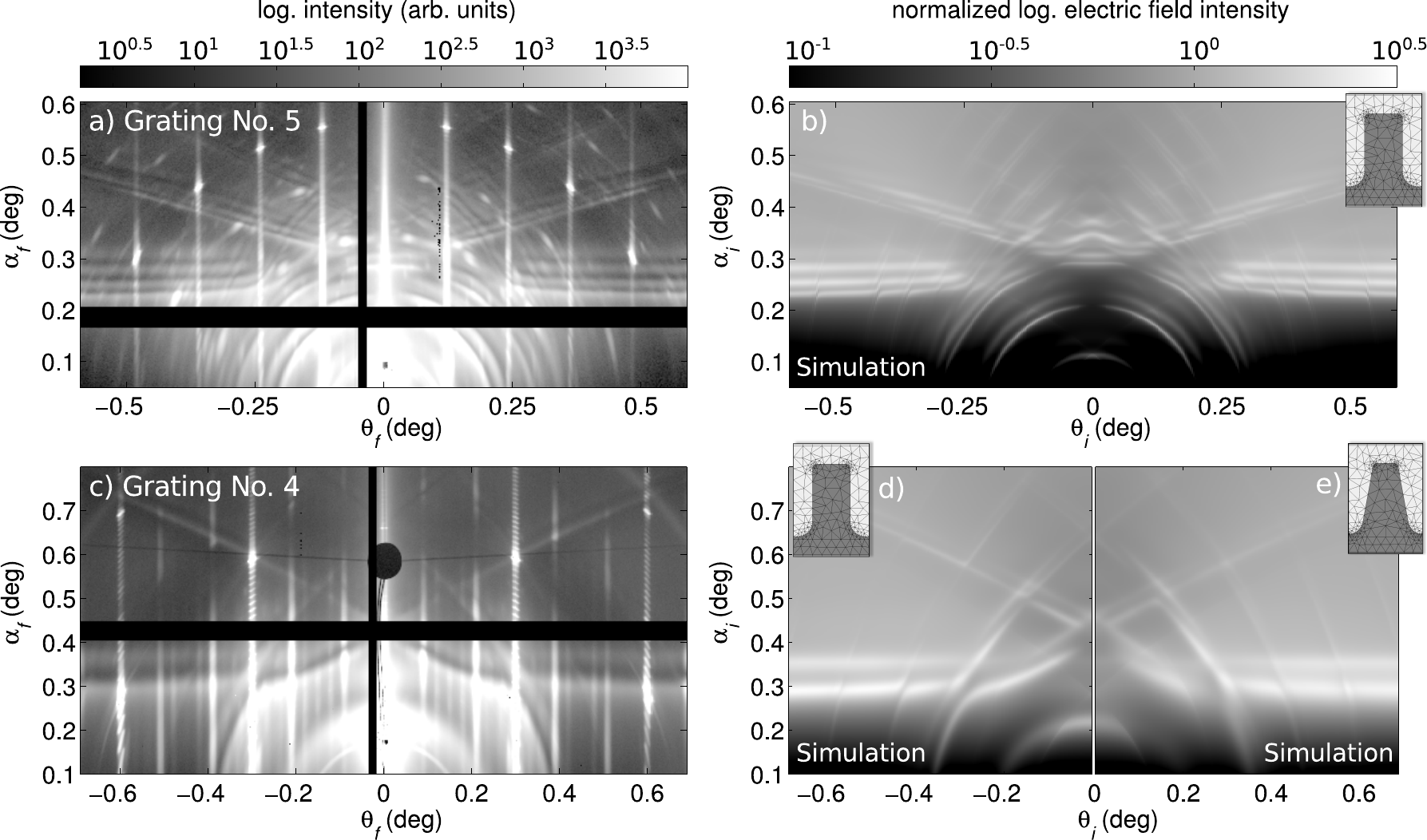}
\caption{Magnified view of the diffuse scattering pattern obtained from (a) grating  No.~5  ($E=6$ keV, $\alpha_i=0.1^\circ$, $\varphi=0^\circ$) and (c) grating No.~4 ($E=5$ keV, $\alpha_i=0.15^\circ$, $\varphi=0^\circ$) around the higher-order Yoneda bands. (b,d,e): Computation of the electric field intensities for both gratings (see text). 
The insets show the rectangular grating model including corner rounding (b,d) and a trapezoidal model with $80^{\circ}$ sidewall angle (e).}
\label{figure_explain_gisaxs3}
\end{figure*}
In a first step the Yoneda effect enhances the diffuse scattering intensity at exit angles $\alpha_f$ close to the critical angle and for arbitrary azimuthal exit angles. This is the 0$^{th}$ order Yoneda band. For a periodic structure like a grating, also higher ordering occurs if the Yoneda-scattered beam is diffracted in a second step at the grating. Clear evidence for the role of the grating diffraction in the higher order Yoneda bands is the fact that an azimuthal rotation of the grating leads to a corresponding shift of these bands preserving the scattering direction with respect to the grating lines (see Fig. \ref{figure_explain_gisaxs2}a).
We calculated the reciprocal space maps for radiation first scattered at $\alpha_f = \alpha_c$, the critical angle of either the substrate or the effective reduced density surface layer, and a random azimutal angle $\theta_f$ and subsequently diffracted at the grating structure,  following the formalism of Yan et al.~\cite{yan_intersection_2007} for a grating azimuthally rotated. The resulting reciprocal space locations for 81 equally spaced values of $\theta_i$, i.e. $\theta_f$ of the first Yoneda scattering event, in the range $\pm0.4^{\circ}$ agree well with the observed scattering pattern (Fig.~\ref{figure_explain_gisaxs2}a red dots).

An azimuthal rotation of the grating allows to separate the higher-order Yoneda bands from other scatter contributions close to the specular axis.
Closer inspection of Fig.~\ref{figure_explain_gisaxs2}b, where the grating was azimuthally rotated with respect to the beam ($\varphi=0.5^{\circ}$), reveals a complex intensity distribution of the higher-order Yoneda bands. The resonant Yoneda enhancement of diffuse scatter contributions interferes with the grating structure at the critical angles of substrate and effective layer and produce a characteristic scattering pattern which is directly related to the geometrical shape of structure. This relation is also demonstrated in Fig.~\ref{figure_explain_gisaxs3} with a comparison of the diffuse scattering patterns obtained in conical scattering geometry ($\varphi=0^\circ$) from grating No.~4 and 5. The distinct differences in the geometrical dimension of the gratings produce a significant variation in the diffuse scattering patterns. To emphasize the higher-order Yoneda bands the gratings were measured in total reflectance with an incidence angle $\alpha_i$ well below the critical angle of substrate and grating effective layer. 

The extremely large computational domain required to represent roughness in a sensible manner renders a direct modeling of diffuse scatter with a finite-element method impractical. We follow here the same reasoning as for the 0$^{th}$ order Yoneda enhancement, i.e. we suppose that the scatter is intense if the electrical field at the grating surface is high. We thus computed the electric fields at the vacuum interface of the grating lines for various incident angles ($\alpha_i, \theta_i$) for both gratings. In analogy to the Yoneda enhancement, the scatter intensity at the corresponding exit angles ($\alpha_f, \theta_f$) should follow this distribution. The computed pattern for a finite-element model similar to the SEM cross section image of grating No.~5 with pronounced corner roundings inside the grooves as shown in Fig.~\ref{figure_explain_gisaxs3}b is indeed very close to the measured scatter distribution in the higher-order Yoneda bands in Fig.~\ref{figure_explain_gisaxs3}a. The simulation of grating No.~4 (cf.~Fig.~\ref{figure_explain_gisaxs3}c and Fig.~\ref{figure_explain_gisaxs3}d) reflects also nicely the measured scatter distribution. To demonstrate the sensitivity for model variations we introduce a reduced sidewall angle ($80^\circ$) in Fig.~\ref{figure_explain_gisaxs3}e to obtain a trapezoidal grating line shape. The clearly visible difference (cf.~Fig.~\ref{figure_explain_gisaxs3}d and Fig.\ref{figure_explain_gisaxs3}e) offers an opportunity for reconstruction of nm sized surface structures based on the measurements of the diffuse scattering pattern.
\subsection{Resonant diffuse scattering}
Apart from Yoneda resonant scattering, a palm-like structure of stacked sheets around the specular axis (see Fig.~\ref{figure_explain_gisaxs}) is observed. Similar patterns can be seen in the figures of previous studies~\cite{gollmer_fabrication_2014,hlaing_nanoimprint-induced_2011} but they were never discussed nor explained.
In contrast to the higher-order Yoneda band, the position of these sheets is invariant with respect to $\alpha_i$ and $\varphi$ variations.
We attribute these sheets to resonant diffuse scattering (RDS) known from multilayer systems~\cite{holy_nonspecular_1994, kaganer_bragg_1995}. 
There, their periodicity in the direction of momentum transfer perpendicular to the surface, $q_z$, represents the layer thickness. For the line gratings we observe similar behavior with respect to the line height, but with an additional dependence on the $q_y$ momentum transfer, which bends the sheets. Measurements of the diffraction efficiencies for various incident angles show an identical modulation for the different diffraction orders. Computations of diffraction efficiencies and X-ray diffraction experiments for trapezoidal gratings  \cite{baumbach_grazing_1999,hofmann_grazing_2009,hu_small_2004} predict a linear shift of the modulation along $q_z$ with $q_y$, whose slope is directly related to the sidewall angle of the grating structure and valid for trapezoidal gratings\cite{Meier_2012}.
SEM cross-section images (cf.~Fig.~\ref{figure_sketch}a) reveal that the line shape of the gratings investigated here deviates from a perfect rectangular or trapezoidal profile, which explains the palm-like bending of the RDS sheets.
\section{Conclusions}
In conclusion, we have presented a full explanation of the diffuse scatter contributions and indicated approaches for numerical modeling which pave the way to exploit these effects for reconstruction of surface structures. In equivalence to X-ray standing wave effects, known from vertically layered systems, a Yoneda band including waveguide modes forms at a laterally structured surface. Here, the grating structure acts as a layer with reduced density and the resulting waveguide modes provide a direct measure of the structure average density, respectively linewidths for rectangular lines at fixed pitch. Higher order Yoneda bands result from enhanced surface scattering and subsequent diffraction at the grating. Their reciprocal space representation is predicted with an analytic approach and a numerical computation of the electric field at the surface of the nano structure correlating the Yoneda resonant enhancement with the surface figure. The palm-like bending of the resonant diffuse scatter sheets is also linked to the surface figure, in our example the shape of the grating lines.\\

\section*{ACKNOWLEDGMENTS}
We thank the European Metrology Research Programme (EMRP) for financial support
within the Joint Research Project IND17 "Scatterometry". The EMRP 
is jointly funded by the EMRP participating countries within EURAMET and the European Union. Furthermore, we acknowledge support by the Einstein Foundation Berlin through ECMath within subproject SE6. 
\bibliographystyle{apsrev4-1}
\bibliography{reference}
\end{document}